\documentclass[aps, prb, twocolumn, floatfix, superscriptaddress, nofootinbib]{revtex4-2}
\usepackage[T1]{fontenc}
\usepackage{amsmath,amssymb,graphicx,epsfig,xcolor}
\usepackage[colorlinks=true,allcolors=blue]{hyperref}
\setcitestyle{super,sort&compress}
\DeclareUnicodeCharacter{0393}{\ensuremath{\Gamma}}
\DeclareUnicodeCharacter{03C1}{\ensuremath{\rho}}
\DeclareUnicodeCharacter{00D7}{\ensuremath{\times}}
\DeclareUnicodeCharacter{2010}{-}
\DeclareUnicodeCharacter{2192}{\ensuremath{\rightarrow}}
\DeclareUnicodeCharacter{2212}{\ensuremath{-}}

\makeatletter
\newcommand{\includepdffigure}[2][]{%
  \begingroup
  \ifdefined\Gread@transgrouptrue
    \let\Gread@transgrouptrue\Gread@transgroupfalse
  \fi
  \includegraphics[#1]{#2}%
  \endgroup
}
\makeatother
\makeatletter
\renewcommand*\frontmatter@thefootnote{\ifcase\csname c@\@mpfn\endcsname\or\#\or\ensuremath{\ast}\else\@arabic\csname c@\@mpfn\endcsname\fi}
\let\FeSO@linked@frontmatter@footnotemark\frontmatter@footnotemark
\renewcommand*\frontmatter@footnotemark[1]{%
  \begingroup
  \let\hyper@linkstart\@gobbletwo
  \let\hyper@linkend\@empty
  \FeSO@linked@frontmatter@footnotemark{#1}%
  \endgroup
}
\makeatother
\begin{document}
\raggedbottom
\title{The Fe--S--O ternary phase diagram and inner-core formation in Mars}
\author{Xiaowei Wang}
\thanks{These authors contributed equally.}
\affiliation{Department of Physics, Xiamen University, Xiamen 361005, China}
\author{Zepeng Wu}
\thanks{These authors contributed equally.}
\affiliation{Department of Physics, Xiamen University, Xiamen 361005, China}
\author{Chen Gao}
\thanks{These authors contributed equally.}
\affiliation{Department of Physics, Xiamen University, Xiamen 361005, China}
\author{Liangrui Wei}
\affiliation{Department of Physics, Xiamen University, Xiamen 361005, China}
\author{Tianqi Wan}
\affiliation{Department of Physics, Xiamen University, Xiamen 361005, China}
\author{Zhen Zhang}
\affiliation{Department of Physics, Xiamen University, Xiamen 361005, China}
\author{Renata M. Wentzcovitch}
\affiliation{Department of Applied Physics and Applied Mathematics, Columbia University, New York, NY 10027, United States}
\affiliation{Department of Earth and Environmental Sciences, Columbia University, New York, NY 10027, United States}
\affiliation{Lamont--Doherty Earth Observatory, Columbia University, Palisades, NY 10964, United States}
\author{Yang Sun}
\thanks{Email: \href{mailto:yangsun@xmu.edu.cn}{yangsun@xmu.edu.cn}}
\affiliation{Department of Physics, Xiamen University, Xiamen 361005, China}

\begin{abstract}
The composition and structure of a possible solid inner core in Mars remain uncertain. Here we combine free energies calculated using a machine-learning interatomic potential with experimental constraints to construct a ternary Fe--S--O phase diagram under Martian core conditions. The phase diagram predicts that B1-structured FeO crystallizes first across the proposed Martian core compositions, with higher sulfur contents lowering the oxygen threshold for its crystallization. The modeled phase boundaries and seismically inferred liquid-core densities suggest two possible scenarios, with sinking FeO forming an inner core directly or buoyant FeO accumulating beneath the core--mantle boundary before a mixed Fe + FeO inner core could form during further cooling. These results provide a thermodynamic basis for understanding the formation and structure of a possible Martian inner core.
\end{abstract}
\maketitle

Understanding the internal structure of rocky planets is central to constraining their origin, differentiation, and thermal evolution\cite{ref1}. Mars is the second planet for which in situ seismic observations are available, which greatly improved constraints on its deep interior\cite{ref2,ref3,ref4}. The seismic and geodetic studies indicate that Mars contains a large metallic core, although its radius and density remain uncertain. Earlier interpretations favored a fully liquid, Fe-rich core\cite{ref3,ref5,ref6,ref7}. More recent seismic evidence has been reported for a solid inner core approximately 600 km in radius\cite{ref8}.

Mars's core is expected to be dominated by Fe-Ni alloy\cite{ref9}. Relative to pure liquid iron, the Martian core inferred from seismological models exhibits a substantial density deficit, indicating that light elements must play a major role in its composition \cite{ref5}. Among the candidate light elements, sulfur is generally regarded as the principal component because of cosmochemical constraints and the sulfur composition in the Martian mantle\cite{ref10,ref11}. Existing estimates allow sulfur contents of up to about 19 wt.\% in the core\cite{ref10}. Oxygen is also expected to be important because it can partition into molten Fe together with sulfur during core formation, and its abundance in the Martian core may reach several wt.\%\cite{ref10,ref11,ref12}. Recent first-principles constraints on liquid Fe-rich alloys under Martian core conditions further show that S and O are particularly effective in lowering the density of liquid iron\cite{ref13}, reinforcing the need to consider the coupled effects of sulfur and oxygen rather than treating the core as a simple binary system. Thus, the phase relations of Fe-rich alloys in the Martian core are directly relevant to its present-day structure and crystallization history.

\begin{figure*}[tp]
\centering
\includepdffigure[width=0.9\linewidth]{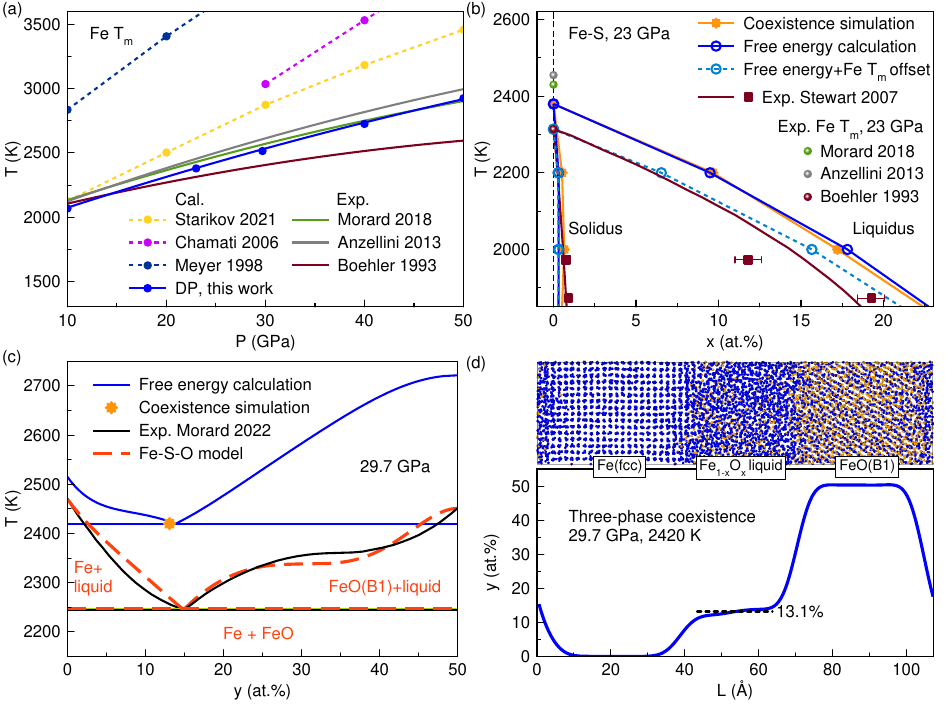}
\caption{\textbf{Phase diagrams of pure Fe and the Fe--S and Fe--O systems.} (a) Melting curve of fcc Fe obtained from SLC simulations using the DP potential, compared with experimental results\cite{ref28,ref29,ref33} and melting curves from SLC simulations using empirical potentials\cite{ref30,ref31,ref32}. (b) Fe--S phase diagram at 23 GPa computed using SLC simulations and free-energy calculations with the DP potential. Brown curves and symbols show the experimental data\cite{ref7}. Sphere symbols denote the experimental melting points of pure Fe\cite{ref28,ref29,ref33}. The dashed curves show the phase diagram recalculated by adding a constant offset, \(\mathrm{\Delta}G_{0} = 6.6\) meV/atom, to the solid-phase free energy to reproduce the pure-Fe melting point used to anchor the experimental phase diagram\cite{ref7}. (c) Fe--O phase diagram at 29.7 GPa computed using free-energy calculations with the DP potential. The orange asterisk marks the eutectic composition determined by the three-phase coexistence simulation at 2420 K. The black curves show the experimental phase diagram\cite{ref34}. The dashed red curves show the phase diagram computed by the Fe--S--O free-energy model. (d) Three-phase coexistence simulation of fcc Fe, Fe--O liquid, and FeO-B1 at 29.7 GPa and 2420 K. The upper panel shows the equilibrium configuration of the coexisting phases. The lower panel shows the oxygen concentration profile perpendicular to the interfaces.}
\label{fig:1}
\end{figure*}

So far, the composition and crystalline structure of the possible Martian solid inner core remain poorly constrained. High-pressure experiments suggested that Fe\textsubscript{12}S\textsubscript{7} can crystallize from Fe--S liquids containing 20--25 wt.\% S at pressures relevant to the center of Mars \cite{ref14}. However, such sulfur-rich compositions are difficult to reconcile with cosmochemical estimates for Mars \cite{ref10}. Other experiments indicate that FeO-B1 can crystallize from Fe--S--O melts containing small amounts of oxygen under Martian core conditions \cite{ref15,ref16}, raising the possibility that oxygen-bearing phases may participate in inner-core formation. The key difficulty is that the relevant phase relations are intrinsically multicomponent, and predicting such a phase diagram generally requires combining experimental phase-boundary constraints with thermodynamic models \cite{ref17}. Due to the lack of multicomponent thermodynamic phase diagrams for Fe alloys under Martian core conditions, the thermodynamic mechanism underlying the formation of a solid inner core remains largely uncertain.

First-principles calculations have provided key constraints on the behavior of materials under high-pressure and high-temperature conditions \cite{ref18,ref19}. Quantitative free-energy calculations and phase-diagram construction generally require extensive sampling of liquid structures, solid-liquid coexistence, and composition-dependent mixing, which remains computationally prohibitive for direct ab initio molecular dynamics (AIMD) \cite{ref20}. Recent advances in machine-learning interatomic potentials now make it possible to retain near-first-principles accuracy while accessing much larger length and time scales \cite{ref21,ref22}, thereby enabling quantitative free-energy calculations and thermodynamic phase-diagram modeling for planetary materials. This approach has begun to address long-standing problems in Earth-core research, including the melting behavior of iron alloys under core conditions \cite{ref23,ref24,ref25,ref26}, and therefore provides a route for extending thermodynamic modeling of Fe alloys under Martian core conditions. Here we develop a machine-learning interatomic potential for Fe--S--O and combine the calculated free energies with experimental constraints on the Fe--S and Fe--O binaries to construct a ternary phase diagram. The model predicts that B1-structured FeO crystallizes first across the proposed Martian core compositions we examine. Increasing the sulfur content lowers the oxygen threshold for FeO to crystallize first. By comparing the density of solid FeO with seismically inferred liquid-core densities, we show how this crystallization sequence could influence the structure of a Martian inner core.

\par\addvspace{6pt}
\noindent\textbf{Deep learning Fe-S-O system}\par\nobreak\vspace{3pt}\makeatletter\@afterindentfalse\@afterheading\makeatother
The machine-learning model is developed to describe interatomic interactions in the ternary Fe--S--O system under Martian core conditions using the Deep Potential (DP) scheme\cite{ref27}. Figure~\ref{fig:1}(a) shows the pure-Fe melting curve from 10 to 50 GPa computed using solid--liquid coexistence (SLC) simulations\cite{ref25}, which agrees closely with recent experimental measurements\cite{ref28,ref29}. For comparison, we also compute melting curves using several previously developed interatomic potentials for iron \cite{ref30,ref31,ref32} and find that these potentials substantially overestimate the melting curve. Among the potentials tested, only the current DP model correctly reproduces the iron melting curve, indicating that it reliably describes the phase competition between liquid and fcc iron under Martian core conditions.

We further compute Fe--S solid--liquid equilibrium using SLC simulations combined with the Monte Carlo (MC) method to accelerate the equilibration of substitutional solid solutions. Figure~\ref{fig:1}(b) shows the resulting Fe--S phase diagram. The calculated solidus agrees well with experimental data\cite{ref7}, while the liquidus is slightly higher. We find that this difference arises because the experimental phase diagram\cite{ref7} is anchored to another pure-Fe melting point data\cite{ref33}, which is 70 K below our prediction. However, as shown in Fig.~\ref{fig:1}(a), the early experimental data\cite{ref33} deviate systematically from more recent measurements \cite{ref28,ref29}. Thus, these comparisons indicate that the DP potential can provide a reasonable description of the interactions in the binary Fe--S system.

Next, we evaluate the ability of the DP model to describe Fe--O interactions by comparing its predictions with the experimental Fe--O phase diagram\cite{ref34} at 29.7 GPa. The free energies of Fe\textsubscript{1-x}O\textsubscript{x} liquids are calculated by thermodynamic integration\cite{ref35} and represented using the quasi-nonideal solution model\cite{ref36,ref37}. The Gibbs free energies of solid fcc Fe and FeO-B1 are determined using their melting points obtained from SLC simulations at 29.7 GPa and extended along the isobar using the Gibbs--Helmholtz relation (Supplementary Note S1). By minimizing the total Gibbs free energy of mixtures of Fe\textsubscript{1-x}O\textsubscript{x} liquid, fcc Fe, and FeO-B1, we obtain the Fe--O solid--liquid phase diagram shown in Fig.~\ref{fig:1}(c). The DP model predicts a eutectic temperature of 2420 K at a composition of 13.9 at.\% O. To validate these results, we perform independent three-phase coexistence simulations of fcc Fe, FeO-B1, and liquid to determine the equilibrium composition and temperature in Fig.~\ref{fig:1}(d). These simulations yield a eutectic liquid composition of 13.1 at.\% O, consistent with the calculated phase diagram. Compared with the experimental phase diagram\cite{ref34}, the DP model overestimates the melting points of pure Fe and FeO-B1 by 50 and 270 K, respectively, and the eutectic temperature by 170 K. Its predicted eutectic composition also deviates slightly from the experimental value of 14.8 at.\% O. Considering that the DP potential provides a reasonably accurate description of the Fe and Fe--S systems, this discrepancy likely arises from the limitations of conventional DFT in capturing electronic correlations in the Fe--O system under Martian core conditions\cite{ref38,ref39,ref40,ref41}.

\par\addvspace{6pt}
\noindent\textbf{Thermodynamic model}\par\nobreak\vspace{3pt}\makeatletter\@afterindentfalse\@afterheading\makeatother
To accurately describe the Fe--S--O system, we construct a thermodynamic model by combining free energies obtained directly from DP calculations with corrections constrained by experimental binary phase-diagram data. We first test the free-energy correction in the Fe--S system at 23 GPa, where the DP coexistence results show a systematic offset from experiment. We compute the Gibbs free energy of the Fe--S liquid and obtain the free energy of the solid solutions using the equilibrium relation\cite{ref42} (Supplementary Note S2). The Fe--S phase diagram is recalculated using a common-tangent construction on the two free-energy curves and agrees with the SLC results (Fig.~\ref{fig:1}(b), Fig. S5(d)). To incorporate experimental constraints, we determine a free-energy offset of \(\mathrm{\Delta}G_{0} = 6.6\ \)meV/atom for solid fcc Fe so that its calculated melting point matches the experimental value of 2314 K\cite{ref7} (Supplementary Fig. S5). We then add this constant offset to all free-energy curves of the Fe\textsubscript{1-x}S\textsubscript{x} solid solution at different temperatures and sulfur contents as

\begin{equation}
G_{\text{Fe-S}}^{\text{Solid, corrected}}(x,T) = G_{\text{Fe-S}}^{\text{Solid}}(x,T) + \mathrm{\Delta}G_{0}.
\label{eq:1}
\end{equation}

By repeating the common-tangent construction, we find that this offset not only reproduces the experimental melting point of pure Fe but also systematically shifts both the solidus and liquidus curves toward the experimental data\cite{ref7} in Fig.~\ref{fig:1}(b). This again indicates that the DP-predicted Fe--S phase diagram is consistent with experiment\cite{ref7}, apart from the difference in the pure-Fe melting-point anchor. Thus, this approach provides an effective way to correct DP free-energy data using experimental constraints.

\begin{figure*}[tp]
\centering
\includepdffigure[width=0.9\linewidth]{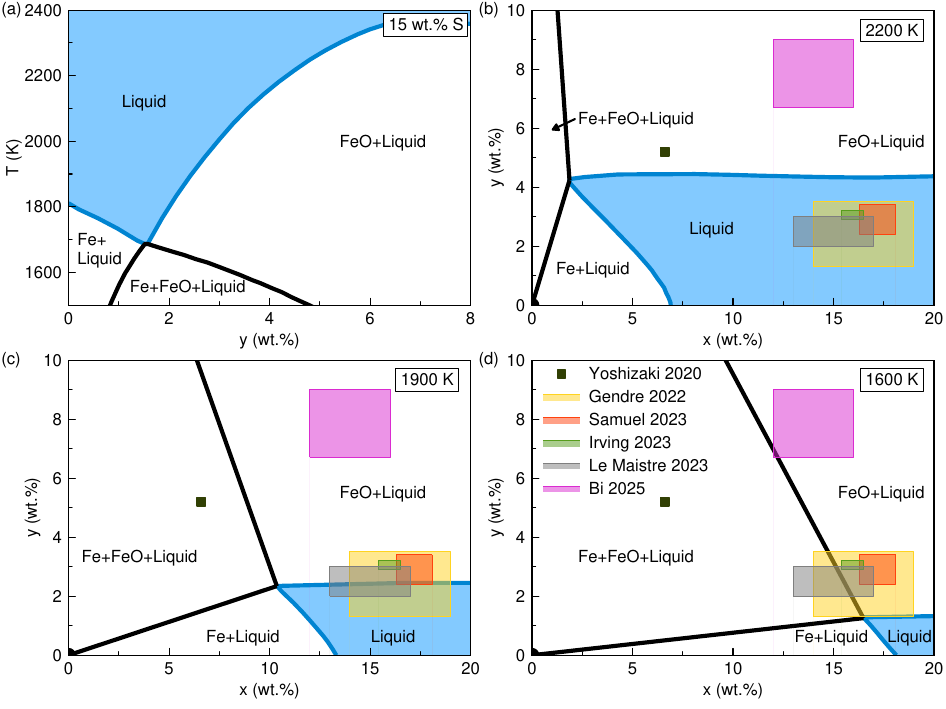}
\caption{\textbf{Fe--S--O ternary phase diagram at 29.7 GPa.} (a) Temperature--oxygen section at a fixed S content of 15 wt.\%. (b--d) Isothermal sections at 2200, 1900, and 1600 K, respectively. Blue shading denotes the Fe--S--O liquid stability field. Colored rectangles and a solid square mark proposed Martian core compositions \cite{ref2,ref3,ref8,ref10,ref44,ref45}.}
\label{fig:2}
\end{figure*}

Next, we use the experimental phase diagram\cite{ref34} to constrain the free energies of the Fe--O system at 29.7 GPa in a similar manner. Temperature-independent offsets of 4.4 and 33.1 meV/atom to the solid--liquid free-energy differences for fcc Fe and FeO-B1 yield melting temperatures of 2471 and 2451 K, respectively, in agreement with the experimental phase diagram (Fig.~\ref{fig:1}(c)). We further apply a composition-dependent free-energy correction, \(\mathrm{\Delta}G_{1}(y)\), to the Fe\textsubscript{1-y}O\textsubscript{y} liquid solution to obtain the corrected free-energy profile

\begin{equation}
G_{\text{Fe-O}}^{\text{L, corrected}}(y,T) = G_{\text{Fe-O}}^{\text{L}}(y,T) + \mathrm{\Delta}G_{1}(y)
\label{eq:2}
\end{equation}

\noindent where \(y\) denotes the oxygen atomic fraction. We determine \(\mathrm{\Delta}G_{1}(y)\) by iteratively fitting the experimental liquidus using the common-tangent equations (Supplementary Note S3). The magnitude of the correction increases nonlinearly with oxygen content (Fig. S2), consistent with the curvature of the experimental liquidus at high oxygen contents. The corrected free energies reproduce the experimental Fe--O phase diagram well without a temperature-dependent term in \(\mathrm{\Delta}G_{1}(y)\), suggesting that the correction is predominantly enthalpic.

As the corrected free energies reproduce the Fe--O phase diagrams well, we extend the thermodynamic model to the ternary Fe--S--O system by incorporating DP free energies of Fe\textsubscript{1-x-y}S\textsubscript{x}O\textsubscript{y} liquids at the same pressure. We apply the same free-energy correction, \(\mathrm{\Delta}G_{1}(y)\), to the ternary liquid. The corrected free energies are parameterized using the quasi-nonideal solution model (Supplementary Notes S1 and S3). We then use the Fe--S--O thermodynamic model to construct the temperature-dependent ternary phase diagram by minimizing the total Gibbs free energy over possible assemblages of fcc Fe, FeO-B1, and Fe\textsubscript{1-x-y}S\textsubscript{x}O\textsubscript{y} liquid (see Supplementary Fig. S7 for the 3D phase diagram and Supplementary Material for an interactive version). We neglect the very limited solubilities of S and O in fcc Fe \cite{ref15}.

Figure~\ref{fig:2}(a) shows a temperature--composition section at 15 wt.\% S, in which FeO-B1 coexists with Fe--S--O liquid over a broad region. Below the liquidus, the fcc Fe+liquid field occurs at oxygen contents below 1.5 wt.\%. At higher oxygen contents, a three-phase field containing fcc Fe, FeO-B1, and Fe--S--O liquid appears. A further increase in oxygen content leads to coexistence between FeO-B1 and Fe--S--O liquid. We also use the 3D phase diagram to examine temperature--composition sections at fixed sulfur contents of 5, 7, and 10 wt.\% (Supplementary Fig. S7). Increasing the sulfur content shifts the junction of the fcc Fe and FeO-B1 liquidus branches toward lower oxygen contents and expands the compositional range of the FeO-B1+liquid field. Thus, higher sulfur contents in the Fe--S--O system reduce the oxygen content required for FeO-B1 crystallization. In experiments at 27 GPa and 1873 K, FeO-B1 coexisted with Fe--S--O liquid at \textasciitilde10 wt.\% S and 2.5 wt.\% O, whereas the equilibrium assemblage at 5.4 wt.\% S and 1.3 wt.\% O consisted of fcc Fe, FeO-B1, and Fe--S--O liquid \cite{ref15}. Pommier et al. \cite{ref43} also observed three-phase coexistence of Fe, FeO, and Fe--S--O liquid at 20 GPa and 1673--1873 K in an Fe--S--O system containing 4.5 wt.\% S and 3 wt.\% O. These observations qualitatively support the predicted stability of FeO-bearing assemblages by our phase diagram.

\begin{figure*}[tp]
\centering
\includepdffigure[width=0.9\linewidth]{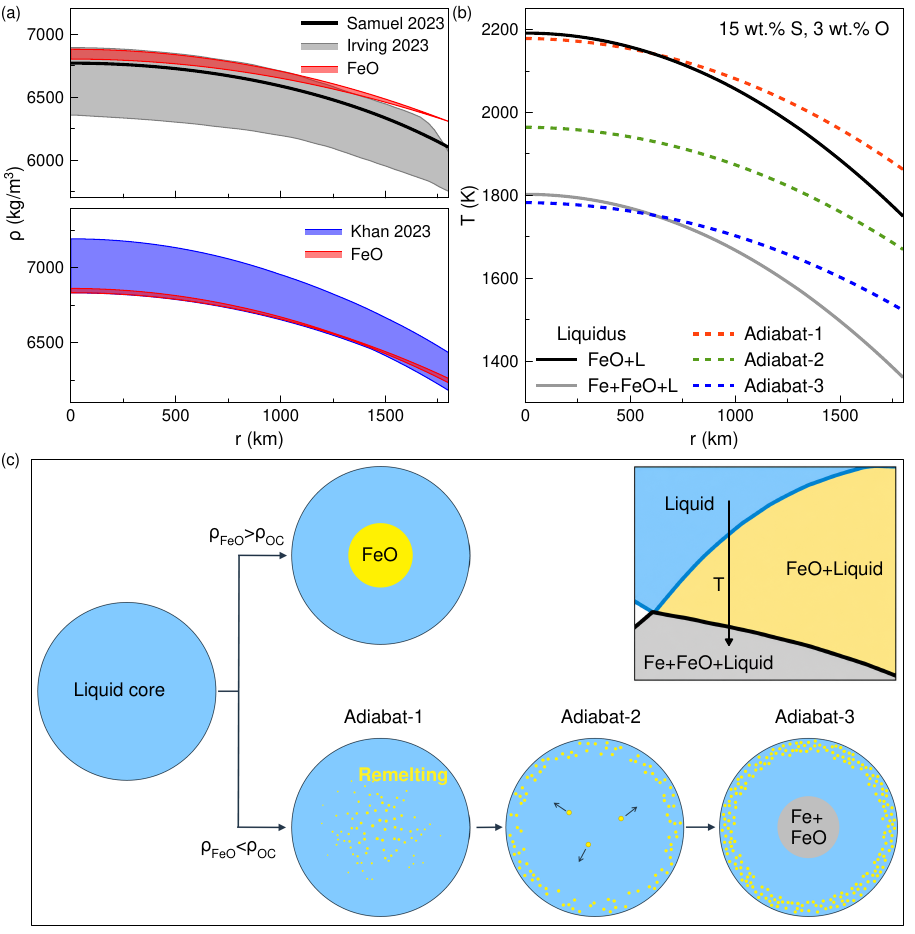}
\caption{\textbf{Implications for Martian core.} (a) Radial density profiles of the Martian core and solid FeO-B1. The upper panel compares the core density profiles of Samuel et al. \cite{ref44} and Irving et al. \cite{ref2} with the corresponding FeO-B1 profiles. The lower panel shows the same comparison for the model of Khan et al. \cite{ref5}. The FeO-B1 profiles (red) are calculated from the experimental pressure--density relation \cite{ref34} using the pressure--radius profile of each model. (b) Temperatures profiles of the two-phase (FeO+liquid) and three-phase (Fe+FeO+liquid) boundaries for a core containing 15 wt.\% S and 3 wt.\% O. The dashed curves show core adiabats at three successive stages of cooling. (c) Schematic evolution of the Martian core in the two scenarios. In Scenario 1 (\(\rho_{\text{FeO}} > \rho_{\text{OC}}\)) FeO crystallizes and accumulates at the center, forming a solid inner core. In Scenario 2 (\(\rho_{\text{FeO}} < \rho_{\text{OC}}\)), newly formed FeO crystals rise and remelt at larger radii (Adiabat-1). With further cooling, the adiabat falls below the FeO+liquid liquidus throughout the core, allowing FeO crystals to rise without remelting and accumulate beneath the CMB (Adiabat-2). Once the adiabat falls below the Fe+FeO+liquid boundary (Adiabat-3), a mixed Fe+FeO inner core begins to form.}
\label{fig:3}
\end{figure*}

We next examine phase relations and crystallization behavior for proposed Martian core compositions using isothermal sections at 2200, 1900, and 1600 K, shown in Figs. 2(b--d), respectively. Square symbols and shaded regions represent previous estimates of the sulfur and oxygen contents of the Martian core \cite{ref2,ref3,ref8,ref10,ref44,ref45}. At 2200 K, the liquid stability field spans a broad compositional range, extending to \textasciitilde4 wt.\% O. The Fe+liquid field occurs at low sulfur and oxygen contents. Most estimated Martian core compositions lie within the liquid stability field at this temperature. Two estimates with relatively high oxygen contents fall within the FeO+liquid field \cite{ref8,ref45}, indicating that FeO-B1 is predicted to have begun crystallizing by 2200 K for these compositions.

As the temperature decreases to 1900 K, the liquidus boundary shifts toward lower oxygen contents, and a larger fraction of the estimated Martian core compositions falls within the FeO+liquid field. For the composition proposed by \cite{ref45}, fcc Fe also becomes stable and coexists with FeO-B1 and Fe--S--O liquid, yielding a mixed solid assemblage of Fe and FeO. At 1600 K, the liquid stability field contracts further, with its upper oxygen boundary decreasing to \textasciitilde1 wt.\% O. At higher oxygen contents, FeO-B1 coexists with Fe--S--O liquid at high sulfur contents and with both fcc Fe and Fe--S--O liquid at lower sulfur contents. Most estimated Martian core compositions therefore fall within the FeO+liquid field, while some extend into the three-phase field. For all the estimated compositions considered here, FeO-B1 is predicted to crystallize first from the Fe--S--O liquid, potentially followed by the formation of Fe+FeO aggregates.

\par\addvspace{6pt}
\noindent\textbf{Martian inner core formation}\par\nobreak\vspace{3pt}\makeatletter\@afterindentfalse\@afterheading\makeatother
For the Martian core compositions considered here, the Fe--S--O phase diagram predicts FeO as the first phase to crystallize. For a core containing 15 wt.\% S and 3 wt.\% O, FeO begins to crystallize at 2005 K at 29.7 GPa (Fig.~\ref{fig:2}(a)). On further cooling to approximately 1616 K, Fe joins the FeO + liquid assemblage. We use this crystallization sequence and the density contrast between FeO and the liquid core to examine how a Martian inner core might form.

Seismic studies infer different pressure--radius profiles for the Martian core \cite{ref2,ref5,ref44} (Supplementary Fig. S8). To compare the density of solid FeO with that of the liquid core, we evaluate the experimentally constrained equation of state for solid FeO \cite{ref34} along the pressure--radius profile of each seismic model. The resulting FeO density profiles differ among the models (Fig.~\ref{fig:3}(a)). In the model by Samuel et al. \cite{ref44}, FeO has a density close to that of the surrounding liquid. Its density lies near the upper bound of the range inferred by Irving et al. \cite{ref2}. Thus, both models allow for a small positive density contrast that could favor central accumulation. In the model of Khan et al. \cite{ref5}, the FeO density lies near the lower bound of the inferred liquid density range throughout the core, so FeO crystals could be buoyant. We therefore consider two scenarios in which newly formed FeO is either denser or less dense than the surrounding liquid.

Figure~\ref{fig:3}(b) compares the radial profiles of the phase boundaries with the core adiabats. We extrapolate the onset liquidus of FeO crystallization at 2005 K and the onset of Fe+FeO+liquid coexistence at 1616 K from 29.7 GPa across the core, using the Clapeyron slope of pure Fe to approximate the pressure dependence of both boundaries. The adiabatic temperature profile \cite{ref46,ref47} of the liquid core is computed as \(T(r) = T_{L}\left( r_{\text{ICB}} \right)\left\lbrack \frac{\rho_{L}(r)}{\rho_{L}\left( r_{\text{ICB}} \right)} \right\rbrack^{1.5}\), where \(T_{L}\left( r_{\text{ICB}} \right)\) is the liquidus at the inner-core boundary (ICB) and \(\rho_{L}(r)\) is the liquid core density at radius \(r\). We use these profiles to explore two crystallization scenarios in Fig.~\ref{fig:3}(c). In Scenario 1, where \(\rho_{FeO} > \rho_{OC}\), FeO begins to crystallize at the center and remains there, allowing the inner core to grow outward. This growth resembles inner-core crystallization on Earth \cite{ref48,ref49,ref47}. Adopting the inner core radius of 613 km inferred by Bi et al. \cite{ref8}, we estimate an effective latent heat of 1200 kJ/kg of FeO formed from the enthalpy difference between the initial liquid and the differentiated state (see Methods). This value is substantially higher than the 750 kJ/kg adopted in the pure-iron inner core model \cite{ref50}. The larger heat release from FeO inner core might require a change of the current Martian thermal and dynamo models.

In Scenario 2, where \(\rho_{FeO} < \rho_{OC}\), Martian inner core crystallization is more complex. When the adiabat cools enough to cross the FeO+liquid liquidus, newly formed FeO crystals rise. Along Adiabat 1 in Fig.~\ref{fig:3}(b), the rising crystals would remelt at larger radii. As the core continues to cool, the adiabat can fall below the FeO + liquid liquidus throughout the liquid core, as illustrated by Adiabat 2 in Fig.~\ref{fig:3}(b). FeO crystals can then rise without remelting and accumulate beneath the core--mantle boundary (CMB). During this stage, however, a solid inner core still cannot form. Only when the adiabat cools enough to cross the Fe+FeO+liquid liquidus (Adiabat 3) does the crystallized assemblage change to Fe+FeO solid aggregates. If the aggregates contain enough metallic Fe to be denser than the surrounding liquid, they can accumulate at the center and form a mixed Fe + FeO inner core. This scenario also offers a possible origin for the solid inner core inferred by Bi et al. \cite{ref8}.

These results show how the density contrast with the liquid can lead to different outcomes for the same crystallization sequence. Improved seismic constraints on the liquid core density profile would help distinguish between them. Scenario 2 also predicts possible FeO enrichment immediately below the CMB. Seismic studies have inferred a basal molten layer above the CMB \cite{ref5,ref44}, and recent studies have examined how FeO enrichment affects the properties and stability of such a layer\cite{ref51,ref52}. Whether an FeO-enriched region below the CMB would produce a distinguishable seismic signature remains to be tested.

To conclude, we constructed a ternary Fe--S--O phase diagram under Martian core conditions to investigate the formation and structure of a possible solid inner core. The thermodynamic model predicts that FeO-B1 crystallizes first and that increasing the sulfur content lowers the oxygen threshold for its crystallization. A central FeO inner core can form if the core adiabat crosses the FeO liquidus near the centre and FeO is denser than the surrounding liquid. In the scenario with buoyant FeO, a mixed Fe+FeO inner core can form, provided that the central region cools into the three-phase field and the solid aggregates are dense enough to sink. Further high-pressure experiments could test the predicted phase relations, while tighter seismic constraints on liquid-core density could help distinguish between the two crystallization histories. An FeO-rich inner core in Mars would incorporate oxygen as it grows, whereas Earth’s Fe-rich inner core rejects oxygen into the liquid outer core. This reversal in oxygen partitioning, together with the distinct crystallization energetics and pathways associated with FeO, may be key to understanding the cooling history of the Martian core and the present-day absence of an active global dynamo.

\par\addvspace{6pt}
\noindent\textbf{Methods}\par\nobreak\vspace{3pt}\makeatletter\@afterindentfalse\@afterheading\makeatother
\par\addvspace{6pt}
\noindent\textbf{Machine-learning interatomic potential}\par\nobreak\vspace{3pt}\makeatletter\@afterindentfalse\@afterheading\makeatother
We developed a machine-learning interatomic potential for Fe--S--O using DP method \cite{ref27} for the temperature--pressure range relevant to the Martian core. The training set comprises four components: (1) pure Fe in the liquid and fcc solid phases; (2) Fe--S liquids containing 15 and 30 at.\% S and fcc substitutional solid solutions containing 5 at.\% S; (3) solid FeO-B1, Fe--O liquids containing approximately 15, 30, 50, and 70 at.\% O, and fcc substitutional solid solutions containing 5 at.\% O; and (4) Fe--S--O liquids with compositions of 7.5 at.\% S + 7.5 at.\% O, 15 at.\% S + 15 at.\% O, 30 at.\% S + 15 at.\% O, and 22.5 at.\% S + 7.5 at.\% O. These sampled Fe--S--O compositions fully encompass the compositional range of the Martian core estimated by Samuel et al. \cite{ref44}.

The dataset was generated in two steps. First, ab initio molecular dynamics (AIMD) simulations combined with the on-the-fly machine-learned force field \cite{ref21} implemented in VASP were performed to collect atomic configurations over -20--90 GPa and 1000--3000 K. Second, DPGEN-based sampling \cite{ref53,ref54} was performed for all compositions within the same pressure--temperature range. The final dataset contains 16,239 configurations listed in Table S6.

DP model training was performed using DeepMD-kit \cite{ref27,ref55}. The neighbor cutoff radius was set to 6.0 Å. The embedding and fitting networks each contained three hidden layers, with \{25, 50, 100\} and \{240, 240, 240\} neurons, respectively. Training was carried out for 1,000,000 steps. The energy prefactor was fixed at 10, whereas the force prefactor was annealed from 1000 to 2 and the virial prefactor from 10 to 1. Electronic entropy, which is essential for describing high-temperature free energy, was accounted in the training \cite{ref25,ref56}. The training results shown in Fig. S9 indicate excellent agreement between DP and DFT calculations, with RMSEs of 5.9 meV/atom for energies, 0.30 eV/Å for forces and 0.27 GPa for pressures.

\par\addvspace{6pt}
\noindent\textbf{Ab initio calculations}\par\nobreak\vspace{3pt}\makeatletter\@afterindentfalse\@afterheading\makeatother
Density functional theory (DFT) calculations were performed using the VASP package \cite{ref57,ref58} with the PAW method \cite{ref59} and the PBE exchange--correlation functional \cite{ref60}. PAW potentials with 8 (3d\textsuperscript{7}4s\textsuperscript{1}), 6 (3s\textsuperscript{2}3p\textsuperscript{4}) and 6 (2s\textsuperscript{2}2p\textsuperscript{4}) valence electrons were used for Fe, S, and O, respectively. The plane-wave cutoff energy was set to 400 eV and Brillouin-zone sampling used Γ point for 216, 250, and 256 atoms. Electronic temperature effects were included using Mermin functional \cite{ref61,ref62}. FeO-B1 was treated in an antiferromagnetic configuration\cite{ref38}, whereas all other structures were calculated without spin polarization.

\par\addvspace{6pt}
\noindent\textbf{Molecular dynamics simulations}\par\nobreak\vspace{3pt}\makeatletter\@afterindentfalse\@afterheading\makeatother
The LAMMPS software package \cite{ref63} were used to perform MD simulations with the DP potential developed in this work. The time step was set to 1 \emph{fs.} The temperature and pressure were controlled using the Nosé--Hoover thermostat and barostat \cite{ref64,ref65}. The SLC method\cite{ref25} was employed to calculate the melting points of fcc-Fe and FeO-B1. The free energy of the liquid was calculated using nonequilibrium thermodynamic integration (NETI) implemented in LAMMPS \cite{ref35}. The Uhlenbeck-Ford model was chosen as the reference state\cite{ref66}.

For the Fe--S binary system, we combined the SLC simulation with the MC method \cite{ref67} to compute the phase diagram between Fe\textsubscript{1-x}S\textsubscript{x} solid and liquid solutions at 1800, 2000, and 2200 K. We constructed an SLC model containing 8192 Fe atoms, with half solid and half liquid. Then, a proportion of Fe atoms in the liquid was randomly replaced by S atoms. During the SLC+MC simulation, 100 trial Fe/S atom swaps were attempted every 100 timesteps and were accepted according to the Metropolis criterion. The total simulation time was 4 ns, which was tested to be sufficiently long to reach equilibrium between the solid and liquid solutions. The equilibrium configuration is shown in Supplementary Fig. S10, where the concentrations of S atoms in the solid and liquid phases are used to represent the solidus and liquidus compositions, respectively.

For the Fe--O binary system, we prepared a three-phase cell containing fcc Fe, Fe--O liquid, and FeO-B1. Starting from an SLC model containing 8192 Fe atoms, we randomly replaced half of the Fe atoms in the liquid region with O atoms. The fcc Fe region was held fixed while the liquid was cooled until FeO-B1 crystallized, producing an fcc Fe--FeO-B1 coexistence structure. The cell was then divided into three equal segments along the \emph{x} direction. We fixed the fcc Fe in the first segment and FeO-B1 in the third segment, replaced all O atoms in the middle segment with Fe atoms, and heated the middle segment to form liquid Fe. A specified fraction of the Fe atoms in the liquid was then replaced with O to obtain the three-phase starting configuration. Finally, the full cell was simulated in the NPT ensemble at 29.7 GPa and 2420 K for 4 ns. The concentration of O atoms in the liquid phases were then determined to obtain the eutectic composition.

\par\addvspace{6pt}
\noindent\textbf{Density profiles under Martian core conditions}\par\nobreak\vspace{3pt}\makeatletter\@afterindentfalse\@afterheading\makeatother
The density of FeO-B1 relative to the surrounding liquid determines whether crystals tend to sink or rise within the Martian core. According to the seismically constrained core density models\cite{ref2,ref5,ref44}, and assuming spherical symmetry, the enclosed mass and gravitational acceleration are calculated from each density profile as

\begin{equation}
m(r) = 4\pi\int_{0}^{r}{\rho(s)s^{2}ds},
\label{eq:3}
\end{equation}

\begin{equation}
g(r) = \frac{Gm(r)}{r^{2}},
\label{eq:4}
\end{equation}

\noindent where m(r) is the mass enclosed within r, ρ(r) and g(r) are the density and gravitational acceleration at radius r, respectively, and G is the gravitational constant. Integration of the hydrostatic equilibrium equation gives the pressure profile

\begin{equation}
P(r) = P_{\text{CMB}} + \int_{r}^{R_{\text{CMB}}}{\rho(s)g(s)ds},
\label{eq:5}
\end{equation}

\noindent where \(P_{\text{CMB}}\) is the pressure at the CMB, and \(R_{\text{CMB}}\) is the core radius for the corresponding model \cite{ref2,ref5,ref44}. The resulting pressure profiles are shown in Fig. S8. Because the temperature dependence of FeO-B1 density is weak\cite{ref34,ref68,ref69}, we evaluated its density along each pressure profile using the 2000 K equation of state of Morard et al. \cite{ref34}.

\par\addvspace{6pt}
\noindent\textbf{Latent heat of FeO crystallization}\par\nobreak\vspace{3pt}\makeatletter\@afterindentfalse\@afterheading\makeatother
In Scenario 1, we considered a present-day solid FeO-B1 inner core with a radius of 613 km \cite{ref8}, surrounded by an Fe--S--O liquid outer core. For the density model of Irving et al.\cite{ref2}, the red and gray shaded regions in Fig.~\ref{fig:3}(a) represent the density ranges of the FeO-B1 inner core and the liquid outer core, respectively. We obtained inner- and outer-core masses of 6.5×10\textsuperscript{21} kg and 1.4×10\textsuperscript{23} kg, respectively, using the density model from ref. \cite{ref2} and an inner-core radius of 613 km\cite{ref8}. The outer-core composition is modeled using \(\mathrm{Fe}_{1 - x - y}\mathrm{S}_{x}\mathrm{O}_{y}\). The numbers of atoms in the inner and outer cores, \(n_{\text{B1}}\) and \(n_{\text{L}}\), respectively, can be determined from mass conservation \(n = \frac{mN_{A}}{\overline{M}}\), where \(m\) is the mass of the inner or outer core, \(N_{A}\) is Avogadro constant, and \(\overline{M}\) is the mean molar mass. The crystallization reaction can then be written as

\begin{widetext}
\begin{equation}
n_{0}\mathrm{Fe}_{1 - x_{0} - y_{0}}\mathrm{S}_{x_{0}}\mathrm{O}_{y_{0}}(\text{L}) \rightarrow n_{\text{B1}}\mathrm{Fe}_{0.5}\mathrm{O}_{0.5}(\text{B1}) + n_{\text{L}}\mathrm{Fe}_{1 - x - y}\mathrm{S}_{x}\mathrm{O}_{y}(\text{L}),
\label{eq:6}
\end{equation}
\end{widetext}

\noindent where \(n_{0}\) is the total number of atoms in the core, whereas \(x_{0}\) and \(y_{0}\) are the sulfur and oxygen atomic fractions in the precursor liquid. The reaction enthalpy per atom of FeO-B1 is

\begin{equation}
\begin{aligned}
&\mathrm{\Delta}H(x,y,T) = H_{\mathrm{Fe}_{0.5}\mathrm{O}_{0.5}}^{\text{B1}}(T)
\\ &\quad{}+ \frac{n_{\text{L}}}{n_{\text{B1}}}H_{\mathrm{Fe}_{1 - x - y}\mathrm{S}_{x}\mathrm{O}_{y}}^{\text{L}}(T)
\\ &\quad{}- \frac{n_{0}}{n_{\text{B1}}}H_{\mathrm{Fe}_{1 - x_{0} - y_{0}}\mathrm{S}_{x_{0}}\mathrm{O}_{y_{0}}}^{\text{L}}(T),
\end{aligned}
\label{eq:7}
\end{equation}

\noindent where the enthalpies are calculated using the Fe--S--O free-energy model

\begin{equation}
H(x,y,T) = G(x,y,T) - T\left. \ \frac{\partial G}{\partial T} \right|_{P,composition}.
\label{eq:8}
\end{equation}

We considered two representative Martian core compositions, one containing 15 wt.\% S and 3 wt.\% O\cite{ref2,ref3,ref10,ref44} and the other containing 7 wt.\% S and 5 wt.\% O\cite{ref45}. At 29.7 GPa and 2000 K, the calculated latent heats of crystallization were 1258 and 1220 kJ/kg, respectively.

\par\addvspace{6pt}
\noindent\textbf{Acknowledgments}\par\nobreak\vspace{3pt}\makeatletter\@afterindentfalse\@afterheading\makeatother
Work at Xiamen University was supported by the National Natural Science Foundation of China (Grants T2422016 and 42374108). R.M.W. acknowledges support from the Gordon and Betty Moore Foundation Award GBMF12801 (doi.org/10.37807/GBMF12801). The supercomputing time was supported by the IKKEM Intelligent Computing Center. Some simulations were performed on the Bridges-2 system at PSC, the Anvil system at Purdue University, the Expanse system at SDSC, and the Delta system at NCSA through allocation DMR180081 from the Advanced Cyberinfrastructure Coordination Ecosystem: Services \& Support (ACCESS) program, which is supported by NSF Grants No. 2138259, No. 2138286, No. 2138307, No. 2137603, and No. 2138296.

\end{document}